\documentclass{article}
\usepackage{ams}
\title{ON A TWO-DIMENSIONAL SYMPLECTIC SPACE-TIME}
\author{Joachim Nzotungicimpaye\footnote{Previously professor at the University of Burundi}\\Kigali Institute of Education\\Department of Mathematics\\P.O.Box 5039\\e-mail kimpaye @kie.ac.rw}

\begin{document}
\maketitle
\begin{abstract}
 We realize the hamiltonian action of the one spatial Poincare group $G$,
 analogous of the projective unitary representation, on a coadjoint orbit
 of $G$ which we interpret as a symplectic space-time where the time is
 associated to a physical quantity of cohomological origin (the force $f$).
 This time is the time delay necessary for the system under the force f to have
 momentum $p$.
\end{abstract}
\section{Introduction}
One can verify that the one dimensional Poincare group
\cite{lev}(Galilei group \cite{lev})$G$ is a subgroup of the real
inhomogeneous symplectic group $ISP(2,R)$. It can then be seen as
a group of canonical transformations for a certain two dimensional
symplectic manifold. In this paper , using the methods of
symplectic realizations of Lie groups \cite{gia}, we realize such
a symplectic manifold as a space-time which is a coadjoint orbit
of the Poincare group $G$ on the dual $\cal{H^*}$of the extension
of its Lie algebra. By the contraction process \cite{wig}, we also
obtain the realization of the one spatial Galilei group on the
same space-time. The paper is organized as follows. In section 2
we study the central extension $H$ of $G$ and the coadjont action
of $G$ on $\cal{H^*}$. In the section 3 we realize the one spatial
space-time as an coadjoint orbit of $G$ and obtain the Poincare
and the Galilei groups as canonical transformations of that
space-time.
\section{Central extension of $G$ and its coadjoint action on $\cal{H^*}$}
First of all, let us recall that the general element of the one
spatial Poincare group $G$ i s parametrized as $g=(v,\tau,x)$where
v is a Lorentz boost, $\tau$ is a time translation while x is a
space translation. The multiplication law is \cite{lev}
\begin{equation}
(v,\tau,x)(v^\prime,\tau^\prime,x^\prime)=(\frac{v+v^\prime}{1+\frac{vv^\prime}{c^2}},\gamma
\tau^\prime+\gamma \frac{vx^\prime}{c^2}+\tau,\gamma v
\tau^\prime+\gamma x^\prime+x)
\end{equation}
where
\begin{equation}
\gamma^{-1}=\sqrt{1-\frac{v^2}{c^2}}
\end{equation}
$c$ being the velocity of light in a empty space. If $K, P$ and
$E$ are respectively the left invariant vector fields generating
the boosts, time translations and space translations, then the
Poincare Lie algebra $\cal{G}$ has the structure
\begin{equation}
[K,E]=P , [K,P]=\frac{1}{c^2}E
\end{equation}
One can verify \cite{ham} that the central extension $\cal{H}$ of
this Lie algebra is generated by $K, P, E$ and $F$ such that the
nontrivial Lie brackets are
\begin{equation}
[K,E]=P , [K,P]=\frac{1}{c^2}E, [P,E]=F
\end{equation}
Now writing the general element $h$ of the corresponding connected
Lie group $H$ as $h=exp(\zeta F)exp(xP+tE)exp(vK)$ , and using the
Baker-Campbell-Hausdorff formulae, we find that the multiplication
law for H is

$(v,\tau,x,\zeta)(v^\prime,\tau^\prime,x^\prime,\zeta^\prime)$
\begin{equation}
=(\frac{v+v^\prime}{1+\frac{vv^\prime}{c^2}},\gamma
\tau^\prime+\gamma \frac{vx^\prime}{c^2}+\tau,\gamma v
\tau^\prime+\gamma
x^\prime+x,\zeta+\zeta^\prime+\frac{1}{2}\gamma(x-v\tau)\tau^\prime-\frac{1}{2}\gamma(\tau-\frac{vx}{c^2})x^\prime)
\end{equation}
We then verify that the coadjoint action of G on $\cal{H^*}$is\\

$Ad^*_{(v,\tau,x)}(k,e,p,f)$
\begin{equation}\label{eq:coad}
=(k+\gamma(x-v\tau)\frac{e}{c^2}+\gamma(\tau-\frac{vx}{c^2})p+\frac{1}{2}(\tau^2-\frac{x^2}{c^2})f,\gamma
e-\gamma pv-fx,-\frac{\gamma v}{c^2}e)+\gamma p+ f\tau,f)
\end{equation}
\section{Symplectic structure on space-time}
Using (4), one finds that the Kirillov form \cite{kim}, \cite{gia}
is
\begin{displaymath}
 \left(\begin{array}{ccc}0&-p&-\frac{e}{c^2}\\p&0&f\\\frac{e}{c^2}&-f&0\end{array}\right)\end{displaymath}
and that the coadjoint orbit is characterized by the invariants
$f$ and
\begin{equation}
{\cal K}=k+f\frac{q^2}{2c^2}-f\frac{t^2}{2}
\end{equation}
where $(t=\frac{p}{f},q=-\frac{e}{f})$. We denote if by $ {\cal
O}_{(f,\cal{K})}$.  The symplectic form is , in Darboux's
coordinates,
\begin{equation}
\sigma=dp\wedge dq
\end{equation}
By use of the result (\ref{eq:coad}),the symplectic realization of
the Poincare group is then given by
\begin{equation}\label{eq:canon}
L_{(v,\tau,x)}(p,q)=(\gamma p+\gamma\frac{fv}{c^2}q+f\tau,\gamma
\frac{v}{f}p+\gamma q+x)
\end{equation}
It depends explicitly on invariant $f$.  This explicit dependence
can be removed if we use , instead of the Darboux's coordiantes
$(p,q)$,  the space-time coordinates $(q,t)$. In that case,
(\ref{eq:canon}) becomes
\begin{equation}\label{eq:spacetime}
L_{(v,\tau,x)}(t,q)=(\gamma t+\gamma\frac{v}{c^2}q+\tau,\gamma v
t+\gamma q+x)
\end{equation}
which are the usual Poincare transformations of space-time. The
dependence on $f$ is transferred to the symplectic form which
become $\sigma$=$fdt\wedge dq$. One can also verify that
\begin{equation}
ds^2=-dq^2+c^2dt^2
\end{equation}
is invariant. The orbit $ {\cal O}_{(f,\cal{K})}$ is then endowed
with two geometric structures: the symplectic one given by (7) and
the pseudo-euclidean one given by (10). If one neglects
$\frac{v}{c}$, the action (\ref{eq:spacetime}) becomes the usual
Galilei transfomation of space-time while (\ref{eq:canon}) becomes
the symplectic realization of the Galilei group on $ {\cal
O}_{(f,\cal{K})}$ \cite{kim}which as we have seen represents a
mssless particle under the constant force $f$. The orbit $ {\cal
O}_{(f,\cal{K})}$ can then be interpreted as a Poincare (Galilei)
elementary system, characterized by a position $q$,  a linear
momentum $p$ under an invariant force $f$. The time $t$ represents
the time delay necessary to the system under the force $f$ to have
momentum $p$.

\end{document}